\documentclass[9pt]{book}

\usepackage[dvips]{graphicx,color}
\usepackage{makeidx,universe}

\makeauthorindex
\BookTitle{The proceedings of the Physics of Accreting
Compact Binaries}

\CopyRight{\copyright 2010 by Universal Academy Press, Inc.}

\begin{document} %%%%%%%%%%%%%%%

%\tableofcontents
\pagenumbering{arabic}

\chapter{The Stunted Outbursts of UU Aquarii \\
Are Likely Mass-transfer Events}

\author{\raggedright \baselineskip=10pt%
{\bf R.\ Baptista,$^{1}$ A.\ Bortoletto,$^{2}$
and R.\ K.\ Honeycutt$^{3}$}\\ %%%%%% <== Authors
{\small \it %
(1) Departamento de F\'{\i}sica, UFSC, Campus Trindade, 
    Florian\'opolis, 88040-900, Brazil \\
(2) Laborat\'orio Nacional de Astrof\'{\i}sica - LNA, Itajub\'a,
    37504-364, Brazil \\
(3) Astronomy Department, Indiana University, Bloomington, IN 
    47405, USA
}
}

%**************************
% Please note:
% One \AuthorContents{} is necessary
%    for EACH CONTRIBUTION, for the contents page and
% One \AuthorIndex{} is necessary
%    for EACH AUTHOR, for the index.
%**************************

\AuthorContents{R. Baptista, A. Bortoletto, and R. K. Honeycutt} 

\AuthorIndex{Baptista}{R.} 
\AuthorIndex{Bortoletto}{A.} 
\AuthorIndex{Honeycutt}{R. K.}

     \baselineskip=10pt
     \parindent=10pt

\section*{Abstract} %%%%%%%%%%%%%%%

We report a time-lapse eclipse mapping analysis of B-band
time-series of UU~Aqr along a typical stunted outburst in
2002 August. Disc asymmetries rotating in the prograde sense
in the eclipse maps are interpreted as a precessing elliptical
disc with enhanced emission at periastron. 
From the disc expansion velocity an $\alpha_{\rm hot}= 0.2$
is inferred. The outburst starts with a 10-fold increase in
uneclipsed light, probably arising in an enhanced disc wind;
the disc response is delayed by 2\,d. The results are 
inconsistent with the disc instability model and suggest that
the stunted outburst of UU~Aqr are the response of its viscous
accretion disc to enhanced mass-transfer events.

\section{Introduction} %%%%%%%%%%%%%%%

UU~Aqr is a 3.9\,h period deeply eclipsing SW~Sex novalike with 
a bright, viscous disc accreting matter onto a $0.67\,M_\odot$
white dwarf (WD) at a rate of $10^{-9}\,M_\odot yr^{-1}$ 
\cite{refbapuu.3}. 
Spectral eclipse mapping reveals an uneclipsed emission line
spectrum responsible for $\simeq 6$\% of the total B-band flux,
probably arising from a vertically-extended chromosphere + disc
wind \cite{refbapuu.2}. With a mass ratio $q = 0.3$, the primary
Roche lobe of UU~Aqr is large enough to allow the accreting matter
to expand beyond the 3:1 resonance radius giving rise to an
elliptical precessing disc and to superhumps in its light curves
\cite{refbapuu.5}. It also shows 0.3~mag brightness modulations
on timescales of a few years (probably caused by 20-50\% long-term
changes in mass transfer rate \cite{refbapuu.1}) as well as 
recurrent 'stunted' outbursts of $\sim 0.6$~mag amplitude which
last for $\sim 5$-7\,d \cite{refbapuu.4}
(suggested to be caused by thermal-viscous instabilities in
its outer and cooler disc regions).

\section{Observations and data analysis} %%%%%%%%%%%

Time-series of B-band CCD photometry of UU Aqr were obtained at
the 0.6\,m telescope of OPD/LNA, in Brazil, during its 2002 August
outburst (started by MJD 54293). The observed outburst was the
second of a series of 4 consecutive stunted-outbursts recurring
at a typical timescale of a month.

Data reduction, light curve extraction and flux calibration
procedures are the same as in Baptista \& Bortoletto 
\cite{refbapuu.1}. The data comprise 6 light curves collected 
along 5 nights, including 4 eclipses. The first two runs frame
the rise to outburst maximum but do not cover the eclipse; two
consecutive eclipses were observed at outburst maximum, and two
others along the declining stages. 
The depth of the eclipse at outburst maximum is the same as in 
quiescence, indicating that the extra light is not from the
eclipsed accretion disc. Moreover, there is little change in the
depth of the eclipse along the decline, indicating that the
accretion disc is not the dominant contributor to the observed
brightness changes.

The light curves were analyzed with eclipse mapping techniques to
solve for a map of the disc surface brightness and for the flux
of an additional uneclipsed component in each case.

\section{Results and discussion} %%%%%%%%%%%%%%%

Light curves and corresponding eclipse maps are shown in Fig.\,1. 
\begin{figure}%%%%%%%%%%%%%%% FIGURE 1
 \begin{center}
   \includegraphics[width=.95\textwidth]{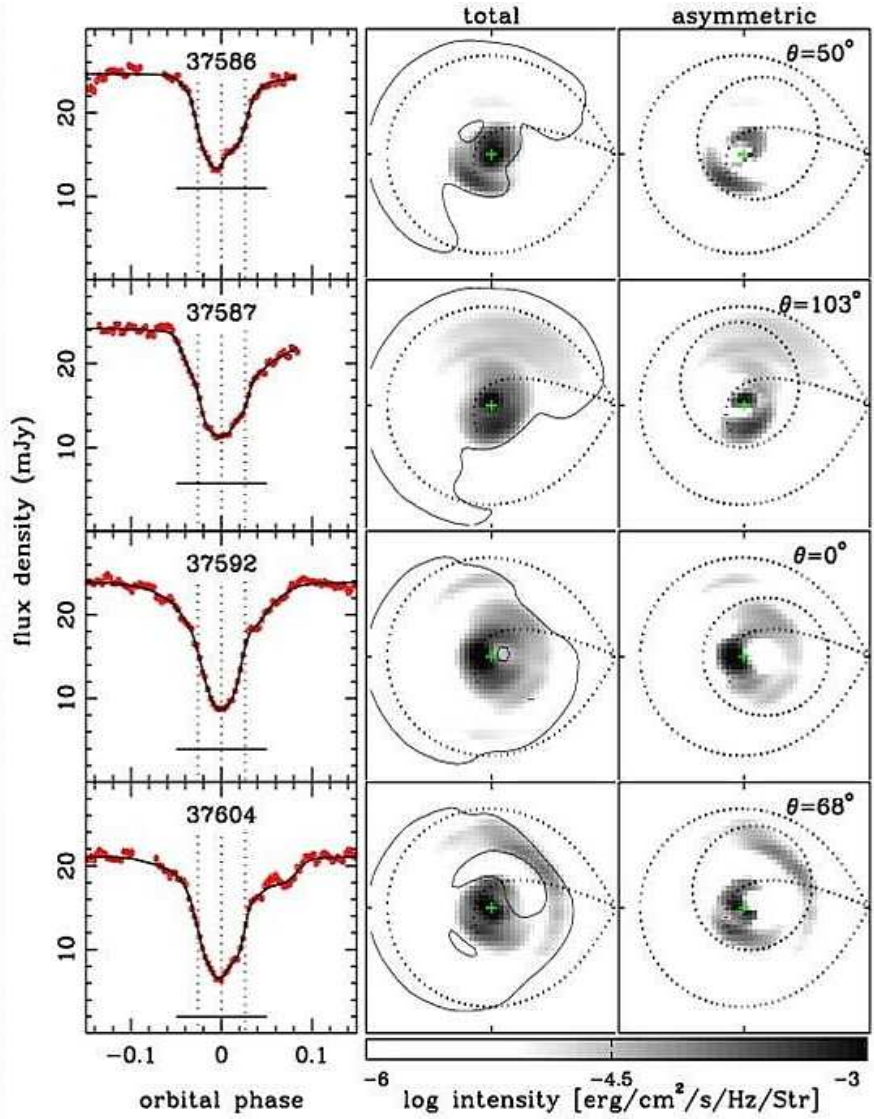}
   \caption{{\bf Left:} Data (red dots) and model (solid lines)
   light curves. Vertical dotted lines mark the ingress/egress
   phases of the WD and mid-eclipse. Horizontal tick marks depict
   the uneclipsed flux in each case and labels indicate the eclipse
   cycle number. {\bf Center:} corresponding eclipse maps in a
   logarithmic grayscale.  Brighter regions are indicated in black;
   fainter regions in white. Dotted lines show the Roche lobe and
   the gas stream trajectory; the secondary is to the right of
   each map and the stars rotate counter-clockwise. A solid contour
   line is overplotted on each map to indicate the 3-$\sigma$
   confidence level region. {\bf Right:}
   The asymmetric component of the eclipse maps in the center
   panels. Additional dotted lines depict the orientation of the
   ellipse of best-fit to the asymmetry in each map. Labels
   indicate the angle between the semi-major axis of the ellipse
   and the line joining both stars. }
    \label{fig:Vienna Univ,}
 \end{center}
\end{figure}%%%%%%%%%%%%%%%
The eclipse maps show an asymmetric brightness distribution with
an arc ellongated in azimuth in the inner regions of the accretion
disc. As the shape of the eclipse changes with time, the resulting
asymmetric arc rotates in azimuth in the prograde sense. Because
the time interval between consecutive eclipse maps is known, it
is possible to measure the rate at which the asymmetry rotates. 
The rotation rate inferred from the first two eclipse maps ($53^o$ 
change after 3.9\,h) is consistent with the orientation of the
asymmetry in the following nights and leads to a precession period
of $P_p= (1.14\pm 0.04)$\,d.

The observed asymmetry cannot be a blob of gas rotating in a
Keplerian orbit around the WD; at that distance from the WD the
Keplerian period is only 14 min, and any structure rotating at
that speed would lead to a blurred ring of emission in an eclipse
map of a $\sim 45$ min long eclipse. 
We tentatively interpret the asymmetry as enhanced emission at the
periastron of an elliptical precessing disc. The beat between the
orbital and precessional periods leads to a predicted superhump
period of $P_s= 1.167 P_{\rm orb}$, longer than the superhump
period of 1.175\,d found by Patterson et~al \cite{refbapuu.5}.

Radial intensity and brightness temperature distributions were
computed from the symmetric component of each eclipse map.
Taking the intensity of the outer radius of the disc in quiescence
as a reference intensity level, we find that the disc expands 
towards outburst maximum at a speed $v_{\rm hot}= +2.0\,km\,
s^{-1}$, and shrinks during the decline at a speed $v_{\rm cool}=
-0.16 \,km\,s^{-1}$. From the expansion velocity we infer a
viscosity parameter $\alpha_{\rm hot}= 0.2$. 

Along the outburst, the brightness temperatures in the outer disc
remain below the critical limit $T_{\rm crit}$ expected for an
outbursting disc according to the disc instability model
\cite{refbapuu.6}; 
the major changes in the temperature distribution occur in the 
disc region already hotter than $T_{\rm crit}$. Moreover, there
is an equivalent critical mass accretion rate \.{M}$_{\rm crit}$
above which the disc stays in the high viscosity state and there
is no more room for disc-instabilities to set in. We find that
$\hbox{\.{M}} > \hbox{\.{M}}_{\rm crit}$ holds at every radius
in quiescence and along the outburst.

The upper panel of Fig.\,2 shows the time evolution of the total
flux, the disc flux and the uneclipsed flux along the outburst.
The lower panel shows the changes along the outburst of the
fractional contribution of the asymmetric sources in the eclipse
map and of the uneclipsed light. 
\begin{figure}[t] %%%%%%  FIGURE 2
 \begin{center}
  \includegraphics[height=26pc]{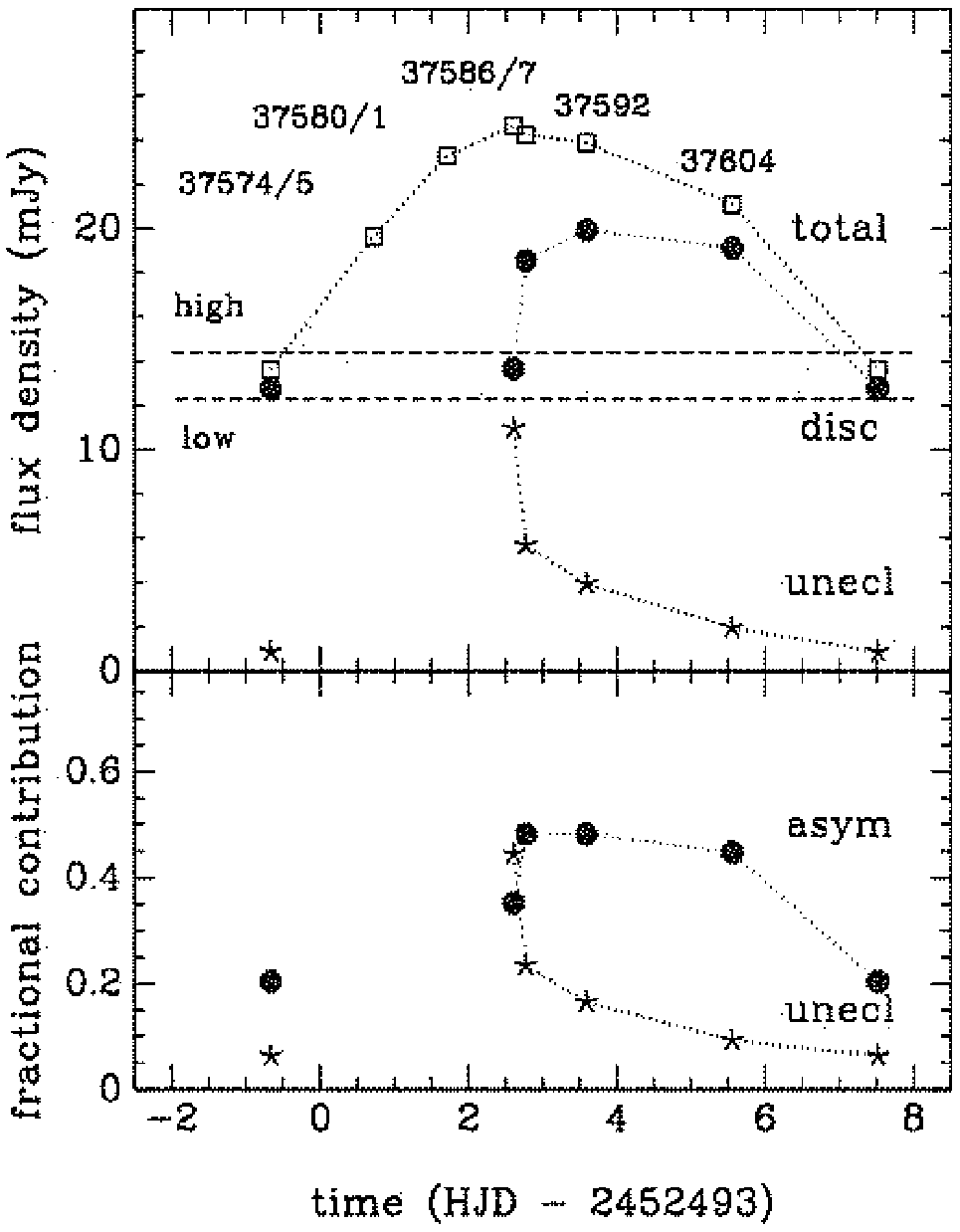}
  \caption{Upper panel: The time evolution of the total flux, the
  disc flux and the uneclipsed flux along the outburst. Average
  values of these quantities for the quiescent state are shown for
  illustration purposes. Labels indicate the eclipse cycle number. 
  Lower panel: the time evolution of the fractional contribution of
  the asymmetric sources in the eclipse map and of the uneclipsed
  light to the total light.}
  \end{center}
\end{figure} %%%%%%%%%%
The rise to outburst maximum is caused by an increase by a factor
10 in the uneclipsed light, which reaches 50\% of the total system
luminosity at outburst maximum and decays exponentially thereafter.
An increase in the luminosity of the mass-donor star by such a factor
would result in conspicuous ellipsoidal modulation and secondary
eclipses, none of which are observed. This suggests that this
uneclipsed light originates in a strongly enhanced disc wind. 
The accretion disc only starts to increase in brightness at
outburst maximum, about two days after outburst onset.

\section{Conclusions} %%%%%%%%%%%%

It is hard to explain the impressive increase in disc wind emission
without any corresponding change in disc brightness in a disc
instability scenario -- where the outburst should start (and be
restricted to) the cooler outer disc regions \cite{refbapuu.6}. 
The combined results suggest that the 'stunted' outbursts of UU~Aqr
are driven by recurrent, fast ($\sim 1\,d$) increases in mass
transfer rate from the donor star by a factor of a few.

%%%%%%%%%%

\end{document}